# Difference-frequency generation with quantum-limited efficiency in triply-resonant nonlinear cavities

Ian B. Burgess,[1,*] Alejandro W. Rodriguez,[2] Murray W. McCutcheon,[1] Jorge Bravo-Abad,[2] Yinan Zhang,[1] Steven G. Johnson,[3] and Marko Lončar[1]

[1]*School of Engineering and Applied Sciences, Harvard University, Cambridge, MA, 02138.*
[2]*Department of Physics, Massachusetts Institute of Technology, Cambridge, MA, 02139.*
[3]*Department of Mathematics, Massachusetts Institute of Technology, Cambridge, MA, 02139.*
*\*Corresponding author: ibburges@fas.harvard.edu*

**Abstract:** We present a comprehensive study of second-order nonlinear difference frequency generation in triply resonant cavities using a theoretical framework based on coupled-mode theory. We show that optimal "quantum-limited" conversion efficiency can be achieved at any pump power when the powers at the pump and idler frequencies satisfy a critical relationship. We demonstrate the existence of a broad parameter range in which all triply-resonant DFG processes exhibit monostable conversion. We also demonstrate the existence of a geometry-dependent bistable region.

**References and links**

1. R.W. Boyd, *Nonlinear Optics,* (Academic Press, CA, 1992).
2. H.M. Gibbs, G. Khitrova, N. Peyghambarian *Nonlinear Photonics*, (Springer-Verlag, 1990).
3. R.A, Baumgartner, R.L. Byer, "Optical Parametric Amplification" IEEE J. Quant. Elec. 15, 432 (1979).
4. J.A. Giordmaine, R.C. Miller, "Tunable Coherent Parametric Oscillation in LiNbO3 at Optical Frequencies" Phys. Rev. Lett. 14, 973 (1965).
5. A. Ashkin, G.D. Boyd, J.M. Dziedzic "Resonant Optical Second Harmonic Generation and Mixing" IEEE J. Quant. Elec. 2, 109 (1966).
6. M.M. Fejer, G.A. Magel, D.H. Jundt, R.L. Byer "Quasi-Phase Matched 2$^{nd}$ Harmonic Generation – Tuning and Tolerances" IEEE J. Quant. Elec. 28, 2631 (1992).
7. L.E. Myers, R.C. Eckardt, M.M. Fejer, R.L. Byer, W.R. Bosenberg, J.W. Pierce "Quasi-Phase Matched Optical Parametric Oscillators in Bulk Periodically Poled LiNbO$_3$" J. Opt. Soc. Am. B 12, 2102 (1995).
8. M. Soljacic, and J. D. Joannopoulos, "Enhancement of nonlinear effects using photonic crystals" Nat. Mater. **3**, 211-219 (2004).
9. M. Bieler, "THz generation from resonant excitation of semiconductor nanostructures: Investigation of second-order nonlinear optical effects" IEEE J. Sel. Top. Quant. Electron. **14**, 458 (2008).
10. A. Alessio, J. Claudon, J.M. Gerard, V. Berger, G. Leo, "Integrated terahertz source based on three-wave mixing of whispering-gallery modes" Opt. Lett. **33**, 2416 (2008).
11. R. E. Haman, M. Ibanescu, E.J. Reed, P. Bernel, S.G. Jonhson, E. Ippen, J.D. Joannopoulos, M. Soljacic, "Purcell effect in nonlinear photonic structures: A coupled-mode theory analysis" Opt. Exp. **16**, 12523 (2008).
12. M.W. McCutcheon, J.F. Young, G.W. Rieger, D. Dalacu, S. Frederick, P.J. Poole, R.L. Williams, "Experimental demonstration of second-order processes in photonic crystal microcavities at submilliwatt excitation powers" Phys. Rev. B. **76**, 245104 (2007).
13. L.-A. Wu, M. Xiao, and H. J. Kimble, "Squeezed states of light from an optical parametric oscillator" J. Opt. Soc. Am. B **4**, 1465–1476 (1987).
14. Z. Y. Ou and H. J. Kimble, "Enhanced conversion efficiency for harmonic generation with double resonance" Opt. Lett. **18**, 1053–1055 (1993).
15. R. Paschotta, K. Fiedler, P. Kurz, and J. Mlynek. "Nonlinear mode coupling in doubly resonant frequency doublers" Phys. Lett. **58**, 117 (1994).
16. V. Berger, "Second harmonic generation in monolithic cavities" J. Opt. Soc. Am. B **14**, 1351 (1997).
17. I. I. Zootoverkh, K. N. V., and E. G. Lariontsev, "Enhancement of the efficiency of second harmonic generation in microlasers" Quant. Elec. **30**, 565 (2000).

18. B. Maes, P. Bienstman, and R. Baets, "Modeling second-harmonic generation by use of mode expansion" J. Opt. Soc. Am. B **22**, 1378 (2005).
19. M. Liscidini and L. A. Andreani, "Second-harmonic generation in doubly resonant microcavities with periodic dielectric mirrors" Phys. Rev. E **73**, 016,613 (2006).
20. Y. Dumeige and P. Feron, "Whispering-gallery mode analysis of phase-matched doubly resonant second-harmonic generation" Phys. Rev. A **74**, 063,804 (2006).
21. A. Rodriguez, M. Soljacic, J. Joannopoulos, S.G. Johnson, "$\chi^{(2)}$ and $\chi^{(3)}$ harmonic generation at a critical power in homogeneous doubly resonant microcavities" Opt. Exp. **15**, 7303 (2007).
22. W. Suh, Z. Wang, S. Fan, "Temporal coupled-mode theory and the presence of non-orthogonal modes in lossless multimode cavities" IEEE J. Quantum Electron **40**, 1511 (2004).
23. J. Bravo-Abad, A. Rodriguez, P. Bernel, S.G. Johnson, J.D. Joannopoulos, M. Soljacic, "Enhanced nonlinear optics in photonic-crystal microcavities" Opt. Exp. **15**, 16161 (2007).
24. H. Hashemi, A.W. Rodriguez, J. D. Joannopoulos, M. Soljacic, and S.G. Johnson. "Nonlinear harmonic generation and devices in doubly resonant Kerr cavities" Phys. Rev. A **79**, 013812, (2009).
25. Y. A. Morozov, I.S. Nefedov, V.Y. Aleshkin, and I.V. Krasnikova, "Terahertz Oscillator Based on Nonlinear Frequency Conversion in a Double Vertical Cavity" Semiconductors **39**, 113 (2005).
26. Y.H. Avetisyan "Cavity-enhanced terahertz region difference frequency generation in surface-emitting geometry" Proc. SPIE 3795, 501.
27. P.B. Deotare, M.W. McCutcheon, I.W. Frank, M. Khan, M. Loncar, arXiv 0901.4158, accepted for publication in Appl. Phys. Lett.
28. S.S. Jha, N. Bloembergen, Physical Review, 171, 891 (1968).
29. S. Saltiel Y.S. Kivshar "Phase matching in nonlinear $\chi^{(2)}$ photonic crystals" Opt. Lett. 25, 1204 (2000).
30. K.L. Vodopyanov, M.M. Fejer, X. Yu, J.S. Harris, Y.-S. Lee, W. C. Hurlbut, V. G. Kozlov, D. Bliss, and C. Lynch, "Terahertz-wave generation in quasi-phase-matched GaAs," Appl. Phys. Lett., vol. 89, pp. 141119, (2006).
31. G. Imeshev, M. E. Fermann, K. L. Vodopyanov, M. M. Fejer, X. Yu, J. S. Harris, D. Bliss, and C. Lynch, "High-power source of THz radiation based on orientation-patterned GaAs pumped by a fiber laser" Opt. Exp. 14, 4439 (2006).
32. J. A. Armstrong, N. loembergen, J. Ducuing, and P. S. Pershan, "Interactions between light waves in a nonlinear dielectric," Phys. Rev., vol. 127, pp. 1918–1939, 1962.
33. F.S. Febler and J.H. Marburger, "Theory of nonresonant multistable optical devices" Appl. Phys. Lett. 28, 731 (1976).
34. S. Schiller, *Principles and Applications of Optical Monolithic Total-Internal-Reflection Resonators*. PhD thesis, Stanford University, Stanford, CA, March 1993.
35. E. Abraham, W. J. Firth, and J. Carr, "Self-oscillation and chaos in nonlinear Fabry-Perot resonators with finite response time," Phys. Lett. A, pp. 47–51, 1982.
36. A. Parini, G. Bellanca, S. Trillo, M. Conforti, A. Locatelli, and C. De Angelis, "Self-pulsing andm bistability in nonlinear Bragg gratings," J. Opt. Soc. Am. B, vol. 24, pp. 2229–2237, 2007.
37. M. Notomi, A. Shinya, S. Mitsugi, G. Kira, E. Kuramochi, and T. Tanabe, "Optical bistable switching action of Si high-q photonic-crystal nanocavities," Opt. Exp., vol. 13, no. 7, pp. 2678–2687, 2005.
38. J. D. Joannopoulos, S.G. Johnson, R. D. Meade, and J. N. Winn, *Photonic Crystals: Molding the Flow of Light* (Princeton Univ. Press, 1995).
39. H. Haus, *Waves and Fields in Optoelectronics* (Prentice-Hall, Englewood Cliffs, NJ, 1984).
40. B.E.A. Saleh, M.C. Teich *Fundamentals of Photonics* (Wiley Interscience, 2007).
41. M.W. McCutcheon, D.E. Chang, Y. Zhang, M.D. Lukin, M. Loncar, "Broad-band spectral control of single photon sources using a nonlinear photonic crystal cavity" manuscript in preparation (2009).
42. M. Tabor, *Chaos and Integrability in Nonlinear Dynamics: An Introduction* (Wiley, New York, 1989).
43. A. Hurwitz, "On the conditions Under Which an Equation Has Only Roots with Negative Real Parts," Mathematicsche Annalen 46, 1895, pp. 273-284. Also in selected papers on Mathematical Trends in Control Theory, Dover, New York, pp. 70-82, 1964.

## 1. Introduction:

Cavity resonant enhancement of nonlinear processes has been studied for several decades, and intra-cavity nonlinear frequency conversion is now used to generate coherent light across much of the electromagnetic spectrum from a small number of primary laser sources [1-7]. In the last decade, there has been renewed interest in applications of nonlinear frequency conversion with the emergence of nanoscale photonic structures which can confine light to ultra-small volumes for long periods of time, allowing for the strong interaction of light with matter [8-29]. For example, nonlinear processes have been considered recently for the generation of coherent radiation in the THz frequency range [9-12,25-26,30-31]. There has been some prior theoretical work on nonlinear frequency conversion in resonant systems [5,9-11,13-21,32-37]; however, these processes are generally treated in the undepleted pump limit, in which reverse processes are neglected, and which lead to incorrect results in the limit of high conversion efficiency. A theoretical framework based on coupled-mode theory has been developed for second ($\chi^{(2)}$) and third ($\chi^{(3)}$) harmonic generation in doubly resonant cavities, which remains valid in the high-efficiency limit and the validity of which is not restricted to a specific geometry [21,23-24]. This type of self-consistent treatment is possible in nonlinear mixing where all relevant modes are resonant (e.g. fundamental and second-harmonic) because the resonant enhancement of the density of optical modes at each frequency causes the nonlinear interaction of these modes to be strongly favored over other processes [21,23-24]. In particular, Ref. [21] proved the existence of a critical power (that depends on the cavity parameters, e.g. frequencies, Quality (Q) factors, mode-overlap) at which 100% conversion is possible, and beyond which efficiency decreases [21]. In addition, in the case of $\chi^{(3)}$ harmonic generation, it was shown that the inclusion of self-phase modulation and cross-phase modulation allow for a rich diversity of dynamics, including limit cycles and multi-stable solutions [24].

In this paper we extend the framework of [21-22,38-39] to the case of difference-frequency generation (DFG) caused by a second-order ($\chi^{(2)}$) nonlinear polarization in a resonant photonic structure. In contrast to Refs. [9,11-12,26], we consider the case where both the input pump (highest frequency) and idler waves are resonant, as well as the generated signal wave at the difference frequency[*]. We show that CW conversion from pump to signal with quantum-limited efficiency (complete depletion of the pump) is possible in such a system for *any* power at the pump frequency, provided there is a critical input power at the idler frequency. The steady-state solutions to the coupled wave equations and their corresponding conversion efficiencies are shown to depend only on universal dimensionless parameters, indicating that their structure is qualitatively the same in all geometries. While the stability of these solutions is the only feature that is geometry-dependent, we demonstrate that in a certain region of the pump-idler power-space, there is only one steady-state solution to the coupled-wave equations that is stable in *all* geometries. Within this range, when the pump-idler power relationship necessary for quantum-limited conversion is satisfied, quantum-limited conversion is the *only stable solution* for all triply-resonant geometries. Outside of this region, we show that bistability exists. We show that the effect of linear losses (e.g. scattering and absorption) results in a rescaling of the conversion efficiency and input powers, with the dynamics remaining qualitatively the same. The maximum efficiency is attained in the over-coupled limit, when the Q-factor is limited by coupling to the input/output port.

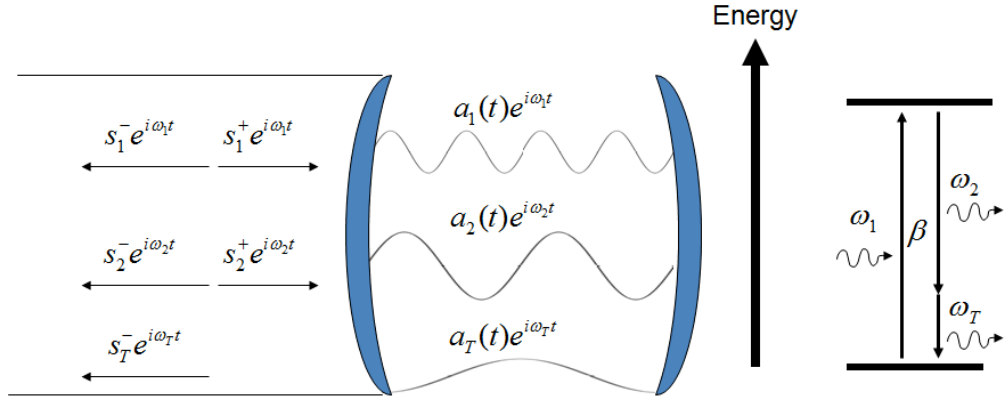

**Figure 1-** Schematic of the CMT framework of triply-resonant DFG. Three resonant modes, $a_k(t)$, are coupled through a second order nonlinear interaction (satisfying $\omega_T = \omega_1 - \omega_2$). All three modes leak energy at a rate, $\gamma_k$, into the outgoing waves, $s_k^-$. The monochromatic input waves which drive the pump ($\omega_1$) and signal ($\omega_2$) modes are represented by the amplitudes, $s_k^+$. The coupling strength is described by the constant, $\beta$.

---

[*] In contrast to the convention used for DFG in optical parametric amplifiers, we refer to the amplified input wave as "idler" and the generated wave as "signal", assuming that it is the signal which is desired for application (e.g. THz generation)

## 2. Coupled Wave Analysis:

The general framework of our temporal CMT model is shown in Figure 1. Our cavity has three resonant modes at frequencies, $\omega_1$ (pump), $\omega_2$ (idler), and $\omega_T$ (signal) which satisfy $\omega_T = \omega_1 - \omega_2$. The electric field in each cavity mode, $E_k(\mathbf{r},t)$, is decomposed into $E_k(\mathbf{r},t) = E_{k,0}(\mathbf{r})a_k(t)\exp(i\omega_k t)$, where $E_{k,0}$ are the normalized mode profiles ($\int d^3\mathbf{r}\varepsilon|E_{k,0}|^2 = 1$) and $a_k(t)$ are the slowly varying wave amplitudes, normalized so that $|a_k|^2$ is the energy stored in each resonant mode [21]. Energy is coupled in/out of each cavity mode through a designated input/output port (e.g. waveguide). The propagating modes in the input/output ports are similarly represented. The relevant incoming and outgoing wave amplitudes are represented by the variables $s_k^+(t)$ and $s_k^-(t)$, respectively, where $|s_k^\pm(t)|^2$ is the power of the propagating mode [21]. Leakage from a cavity mode is described by the loss rate, $\gamma_k = \omega_k/2Q_k$, where $Q_k$ is the Quality factor of the $k^{\text{th}}$ mode. We decompose this cavity loss rate into $\gamma_k = \gamma_{k,s} + \gamma_{k,l}$ [21-22,38-39], where, $\gamma_{k,l}$ represents intrinsic losses (absorption/scattering), while $\gamma_{k,s}$ represents coupling to the input/output port. The incoming and outgoing waves are related by [21,38-39]

$$s_k^- = -s_k^+ + \sqrt{2\gamma_{k,s}}\, a_k \tag{1}$$

A detailed description of the CMT formalism for nonlinear mixing between resonant modes is described in refs. [14, 21-22, 38-39]. The theory assumes that the nonlinear polarizations are sufficiently weak to be treated in the perturbative limit ($\delta P_{k,\text{NL}} \ll P_k$). By extending the formalism in ref. [21] to non-degenerate three-wave mixing, where we assume there is no input at the signal frequency ($s_T^+ = 0$), the coupled wave equations can be written as

$$\frac{da_1}{dt} = -\gamma_1 a_1 - i\omega_1 \beta_1 a_2 a_T + \sqrt{2\gamma_1 \Gamma_1}\, s_1^+ \tag{2}$$

$$\frac{da_2}{dt} = -\gamma_2 a_2 - i\omega_2 \beta_2 a_1 a_T^* + \sqrt{2\gamma_2 \Gamma_2}\, s_2^+ \tag{3}$$

$$\frac{da_T}{dt} = -\gamma_T a_T - i\omega_T \beta_T a_1 a_2^* \tag{4}$$

where $\Gamma_k \equiv \gamma_{k,s}/\gamma_k$ is the fraction of power escaping from the cavity that is collected into the input/output waveguide. The coupling constants ($\beta_k$) are determined by first-order perturbation theory, with the normalization such that $|a_k|^2$ is the energy stored in each mode to remain consistent with Eqs. (1-4) [21-22,38-39]. This yields the following expression for the coupling coefficients:

$$\beta \equiv \beta_1 = \beta_2^* = \beta_T^* = \frac{1}{4}\frac{\int d^3\vec{x}\sum_{i,j,k}\varepsilon_0 \chi_{ijk}^{(2)} E_{1,i}^*(E_{2,j}E_{T,k} + E_{T,j}E_{2,k})}{\sqrt{\int d^3\vec{x}\varepsilon|\vec{E}_1|^2}\sqrt{\int d^3\vec{x}\varepsilon|\vec{E}_2|^2}\sqrt{\int d^3\vec{x}\varepsilon|\vec{E}_T|^2}} \tag{5}$$

where $\varepsilon = \varepsilon_r\varepsilon_0$ is the dielectric constant of the considered material at the corresponding resonance frequency. The above equations (subject to Kleimann symmetry in $\chi^{(2)}$) satisfy the energy conservation condition, $\beta_1 = \beta_2^* = \beta_T^*$. Hereafter we will replace the parameters, $\beta_k$, with the single parameter, $\beta$, which is defined in Eq. (5), with $\beta$ represented in units of $J^{-1/2}$.[**] Optimization of the mode overlap, $\beta$, in Eq. (5) is analogous to satisfying the phase-matching condition in the nonlinear mixing of propagating modes.

## 3. Quantum-Limited Conversion:

DFG efficiency can be assessed by analyzing the steady state solutions ($da_k(t)/dt = ds_k^\pm(t)/dt = 0$) of the coupled-wave equations (2-4). In contrast with degenerate nonlinear frequency conversion (e.g. second-harmonic generation) [21], the non-degenerate case (e.g. DFG) does not allow for 100% conversion of power from $\omega_1$ to $\omega_T$. This arises because the destruction of a photon at $\omega_1$ and the creation of a photon at $\omega_T$ is always accompanied by the creation of a second photon at $\omega_2$. This fundamental constraint is described by the Manley-Rowe equations [1]. Considering also losses ($\Gamma_k$), the quantum limit of conversion efficiency is given by:

$$\frac{|s_T^-|^2}{|s_1^+|^2} \leq \frac{\omega_T}{\omega_1}\Gamma_1\Gamma_T \tag{6}$$

---

[**] Note that there is a typographical error in the corresponding equation for the degenerate case in [21], where $\varepsilon$ is written instead of $\varepsilon_0$ in the numerator.

This condition can be derived directly from the steady-state coupled wave equations, and will be shown once we re-express Eqs. (2-4) in dimensionless form. The conversion efficiency is maximized by maximizing the ratios, $\Gamma_1$ and $\Gamma_T$. This means that over-coupling the cavity modes to the input/output waveguide is essential to achieving high conversion efficiency. In other words, it is more important that the Q-factors be limited by leakage to the desired output port than that they have a higher value but lack a well-defined output channel. This is a well-known condition for low-loss operation in resonant devices [38,39]. However, notice that the loss ratio in the idler, $\Gamma_2$, does not appear in Eq. (6), i.e. it does not affect the overall efficiency. This is because the input wave at $\omega_2$ is not down-converted, but rather amplified; therefore, losses can be compensated by pumping in more power at this frequency. One can find a critical relationship between the input powers, $|s_1^+|^2$ and $|s_2^+|^2$ that allows for maximum conversion efficiency (complete depletion of the pump) by imposing either of the following equivalent constraints: $s_1^- = 0$ or an equality in Eq. (6). This relationship is given by:

$$|s_2^+|^2 = \frac{\omega_2 \left|1 - \frac{4}{\omega_1}|\beta|^2 Q_1 Q_2 Q_T \Gamma_1 |s_1^+|^2\right|^2}{16|\beta|^2 Q_1 Q_2 Q_T \Gamma_2} \tag{7}$$

Notice that the power coupled into the cavity at the idler frequency, $|s_2^+|^2$, depletes the signal ($\omega_T$) when nonlinearly converted (via sum-frequency generation), and produces power at the pump ($\omega_1$). Therefore, in the case in which the pump is completely depleted ($s_1^- = 0$), no net power from the idler frequency is converted and $|s_2^-|^2 = |s_2^+|^2 + |s_1^+|^2\Gamma_1\Gamma_2\omega_2/\omega_1$. The conversion of the total input power, $|s_1^+|^2 + |s_2^+|^2$, is thus maximized in the limit $|s_2^+|^2 \to 0$, $|s_1^+|^2 = \omega_1/(4|\beta|^2 Q_1 Q_2 Q_T \Gamma_1)$. Note, however, that Eqs (2-4) require a non-zero value (this can be arbitrarily small) of $|s_2^+|^2$ for non-zero conversion efficiency. Note that the power at which total conversion is optimal decreases with increasing Q-factors and vice-versa. This means that lower Q-factors are desirable for high power applications. Eq. (7) also allows for quantum-limited conversion for an arbitrarily small value of $|s_1^+|^2$ as $|s_2^+|^2 \to \omega_2/(16|\beta|^2 Q_1 Q_2 Q_T \Gamma_2)$. This is a unique property of the non-degeneracy of DFG, and could prove useful for low power applications [38,40-41]. A similar result was found in the quantum limit for doubly resonant DFG, specifically that a single photon could be converted at high efficiency from $\omega_1$ to $\omega_T$, given an appropriate pump power at $\omega_2$ [41]. Note that in both limits the optimal input power is proportional to $1/Q_1 Q_2 Q_T$. A similar result was found in the degenerate case of second-harmonic generation, where the critical power was shown to scale as $1/Q_1 Q_2^2$ [21]. A related enhancement factor was also reported earlier in the undepleted-pump limit [19]. In the specific case of THz generation through DFG of telecom-band modes ($\omega_1 \approx \omega_2$), assuming the two telecom-band modes have the same mode volume ($V_1 \approx V_2$), then under the conditions of best possible overlap [21], the overlap would scale as $\beta \sim 1/V_T^{1/2}$. This means that in triply resonant THz generation, the mode volumes of the telecom-band cavities do not affect the strength of the nonlinear coupling.

### 4. Stability and Universal Structure

We now analyze the structure of Eqs. (2-4) in detail. We aim to identify dynamics that are qualitatively universal to all triply-resonant DFG processes; i.e., they do not depend explicitly on $\{Q_k,\omega_k,\beta\}$. To do this, it is helpful to redefine our variables $\{a_k, s_k^\pm\}$ in a dimensionless form that removes as many of the parameters, $\{Q_k,\omega_k,\beta\}$, as possible. This is best accomplished by rescaling the variables, $\{t, a_k, s_k^\pm\}$, as follows:

$$T \equiv \gamma_1 t = \frac{\omega_1 t}{2Q_1} \quad (8)$$

$$A_k \equiv 2\sqrt{\frac{Q_1 Q_2 Q_T}{Q_k}}[\beta^* + \delta_{k,T}(-i\beta - \beta^*)]a_k \quad (9)$$

$$S_k \equiv 4\sqrt{\frac{Q_1 Q_2 Q_T \Gamma_k}{\omega_k}}\beta^* s_k^+ \quad (10)$$

where $\delta_{k,T}$ is the Kronecker delta. We will also define a dimensionless parameter that quantifies the DFG efficiency:

$$E_{f\!f} \equiv \frac{\omega_1}{\Gamma_1 \Gamma_T \omega_T}\frac{|s_T^-|^2}{|s_1^+|^2} = 4\frac{U_T}{P_1} \quad (11)$$

where we have introduced the notation, $|A_k|^2 \equiv U_k$, $|S_k|^2 \equiv P_k$ for the normalized energy and power, respectively. In this notation, the quantum limit corresponds to $E_{f\!f} = 1$. The dimensionless coupled-mode equations are given by

$$\frac{dA_1}{dT} = -A_1 + A_2 A_T + S_1 \quad (12)$$

$$\frac{dA_2}{dT} = r_2[-A_2 - A_1 A_T^* + S_2] \quad (13)$$

$$\frac{dA_T}{dT} = r_T[-A_T - A_1 A_2^*] \quad (14)$$

where $r_k = \gamma_k/\gamma_1$. Since $S_1$ and $S_2$ denote amplitudes of waves with different frequencies, the time origin can be redefined so that both $S_1$ and $S_2$ are positive real numbers (i.e. the waves are in phase at $T = 0$), assuming that the waves are not phase-modulated. It follows that the normalized mode amplitudes, $A_1$, $A_2$ and $A_T$ are purely real in any steady-state solution. In the steady state, Eqs. (12-14) simplify to,

$$A_1 = \frac{S_1}{(1+U_2)} \quad (15)$$

$$A_2 = \frac{S_2}{(1-U_1)} \quad (16)$$

$$A_T = -A_1 A_2 \quad (17)$$

The steady-state solutions ($dA_k/dT = 0$) do not depend on the geometry-specific parameters, $r_k$, and thus can be treated in a general form for an arbitrary system. Specifically, the conditions for efficient conversion will depend only on the generalized input parameters, $S_k$, whose magnitude is related to the input power by Eq. (10). However, the stability of the steady-state solutions may in general depend on the specific geometry ($r_k$), as shown below. An analogous simplification was also found in third-harmonic generation ($\chi^{(3)}$) in doubly-resonant systems [24]; however, in this case the competition between frequency conversion and self-action processes (self-phase modulation and cross-phase modulation) prevented the structure of the steady-state solutions from being completely geometry-independent. Using Eqs. (15-17), the dimensionless efficiency parameter can be expressed in the steady state purely as a function of the energy in the idler cavity, $U_2$:

$$E_{f\!f} = \frac{4U_2}{(1+U_2)^2} \quad (18)$$

We can now easily verify that the quantum limit is enforced by the coupled-wave equations (2-4); that is, that $E_{ff} \leq 1$ in the steady state. Further, the quantum-limited conversion is achieved when

$$U_2 = 1, U_1 = U_T = \frac{P_1}{4}, P_2 = (1 - \frac{P_1}{4})^2 \tag{19}$$

Stability of steady-state solutions in a nonlinear dynamical system can be analyzed qualitatively based on the eigenvalues of the points' Jacobian matrices. A steady state is stable if all the eigenvalues of its Jacobian have a negative real part [42]. The Jacobian can then be written as:

$$J = \begin{bmatrix} -1 & A_T & A_2 \\ -r_2 A_T & -r_2 & -r_2 A_1 \\ -r_T A_2 & -r_T A_1 & -r_T \end{bmatrix} \tag{20}$$

Figure 2 shows the structure of the steady-state solutions for $r_2 = r_T = 0.4$. The stable solutions with maximum- and minimum-efficiency are shown in Figure 2(a) and 2(b), respectively. The dotted line in Figure 2 indicates a saddle-node bifurcation. In the region to the left of this line (lower values of $P_1$), there is only one steady-state solution. To the right of this line, there are three possible steady states for every combination of CW input powers, two of which are stable.

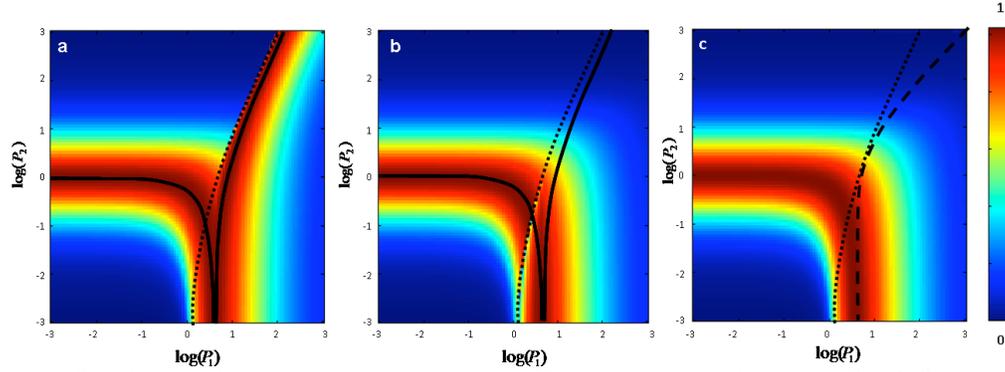

**Figure 2** – Maximum (**a**) and minimum (**b**) normalized efficiency parameter ($E_{ff}$) for stable CW DFG ($E_{ff}$ = 1 corresponds to quantum-limited conversion), plotted as a function of normalized powers of the pump ($P_1$) and idler ($P_2$) input waves. The solid line denotes the critical relationship between $P_1$ and $P_2$ where $E_{ff}$ = 1 is possible ($P_2 = (1 - P_1/4)^2$). The dotted line (shown in both (**a**) and (**b**)) denotes the onset of multi-stability (saddle-node bifurcation). To the left of the dotted line (in all three plots), there is only one steady state solution to the coupled wave equations, which is stable for all geometries (any $r_2$ and $r_T$). To the right of the dotted line, there are three steady-state solutions and bistable behavior is observed for small values of $r_2$ and $r_T$. The two stable solutions are shown in (**a**) and (**b**) (the unstable solution is not shown). Stability in this region was assessed using $r_2 = r_T = 0.4$ (typical values for a near-degenerate coupled cavity system where all the modes have similar Q-factors). The stability of the solutions in the multi-stable region depends on the parameters, $r_2$ and $r_T$ (see Fig. 3 for example). (**c**) Stable conversion efficiency reached after a step-excitation ($U_k(T=0) = 0$, $P_k(T \geq 0)$ = constant, $P_k(T < 0) = 0$) for $r_2 = r_T = 0.4$. To the left of the dashed line, this solution is stable in all geometries (any $r_2, r_T$).

Given that there exists a large region of the $\{P_1, P_2\}$ parameter space with only one solution to Eqs. (15-17), one would expect there to exist a subset of this region where this solution is stable for any geometry $\{r_2, r_T\}$. In fact, it can be shown that *in the entire region this solution is stable in all geometries* (in the region where there is only one solution). This can be proven by analyzing the Jacobian's characteristic polynomial using the Routh-Hurwitz algorithm [43]. The characteristic polynomial of (20) can be written in the form $\lambda^3 + B\lambda^2 + C\lambda + D = 0$. Using Eqs. (15-17) to simplify, we find that:

$$B = 1 + r_2 + r_T \tag{21}$$

$$C = r_2 r_T (1 - U_1) + r_2 (1 + \frac{1}{4} E_{ff} P_1) + r_T (1 + U_2) \tag{22}$$

$$D = r_2 r_T (1 - U_1 + U_2 + \frac{3}{4} E_{ff} P_1) \tag{23}$$

Note that all variables in Eqs. (21-23) can only have real positive values. The signs of the real parts of the Jacobian's eigenvalues can be determined by analyzing the first column of the Routh-Hurwitz matrix. If all entries in the column have the same sign then there are no eigenvalues with positive real parts [43]. For our Jacobian, this column is given by

$$\{H_{i,1}\} = \begin{bmatrix} 1 \\ B \\ \frac{1}{B}(BC - D) \\ D \end{bmatrix} \tag{24}$$

The additional constraint that no eigenvalues have a real part equal to zero imposes the further constraints, $(BC - D), D \neq 0$. If we require $U_1 \leq 1$, all the coefficients (B-D) are strictly positive. In this case, both constraints are reduced to a single condition, $(BC - D) > 0$. The term, $BC - D$, can be expressed as

$$BC - D = r_2 r_T [(r_2 + r_T)(1 - U_1) + 2(1 - \frac{E_{ff} P_1}{4})] + (1 + \frac{E_{ff} P_1}{4})(1 + r_2) r_2 + (1 + U_2)(1 + r_T) r_T \tag{25}$$

Therefore, *a steady-state solution is stable in all geometries when $U_1 < 1$, $P_1 \leq 4/E_{ff}$*. It follows from Eq. (15) that all steady states are stable when $P_1 \leq 1$. In fact, we find numerically that these conditions are satisfied for the steady-state solution everywhere in the region of mono-stability (Fig. 2, to the left of the dotted line). Thus, we expect universal dynamics in this region of $\{P_1, P_2\}$ space, with the conversion efficiency stabilizing to a value determined purely by the input powers, $P_1, P_2$, shown in Fig. 2. Furthermore, for all $P_1 < 4$, the solution with quantum-limited efficiency given in Eq. (19) is universally stable (for all $r_2$ and $r_T$). For Eq. (19) with $P_1 > 4$, stability is guaranteed by (21-25) when $r_T < 0.414$ ($\sqrt{2} - 1$). Figure 3 demonstrates how the stability of the high-efficiency solutions varies with the geometric parameters, $r_2$ and $r_T$, for $P_1 > 4$ when $P_1$ and $P_2$ are related by Eq. (19). When $r_2$ and $r_T$ are both large, there remains only one stable solution (shown by the black curve in Figure 3) for all $P_1$, which has quantum-limited efficiency for $P_1 < 4$.

Stability is also independent of the parameters, $\{r_2, r_T\}$, in the limit $P_2 \ll 1$. We expect there to be multiple steady states for $P_1 > 1$ (see Fig. 2). We can analytically determine the efficiencies and stabilities of all steady-state solutions in this limit. For any value of $P_1 \neq 1$, there is one solution where $U_1 \rightarrow P_1$, $U_2 \rightarrow 0$, and $E_{ff} \rightarrow 0$ (see Eqs. (15-17)). This will be the only solution for $P_1 < 1$. In particular, it follows directly from Eqs. (21-25) that this solution is universally stable for $P_1 < 1$ and universally unstable for $P_1 > 1$. For $P_1 > 1$, there is an additional pair of solutions in which $U_1 \rightarrow 1^+$, and $U_1 \rightarrow 1^-$. These solutions differ only by the sign of $A_2$ (phase), both having $U_2 = P_1^{1/2} - 1$, and $E_{ff} = 4(P_1^{1/2} - 1)/P_1$. It can also be shown directly from Eqs. (21-25) that both of these solutions are stable for all geometries. This means that to a good approximation, all geometries display only one stable conversion efficiency as a function of $P_1$ for $P_2 \ll 1$.

In addition to verifying the stability of high-conversion-efficiency solutions, it is important to assess how these solutions can be excited. In the monostable region, the efficiency will stabilize at the value shown in Fig. 2 for all excitations; however, different excitation patterns can lead to different long-time stabilities in the bistable region. This principle was demonstrated for multi-stable conversion in doubly-resonant third-harmonic generation [24]. Fig. 2(c) shows the asymptotic solution reached after a simple step excitation ($U_k (T = 0) = 0$, $P_k (T \geq 0) = $ constant, $P_k(T < 0) = 0$, $\gamma_2 = \gamma_T = 0.4$). For $P_1 < 4$, quantum-limited conversion is achieved from a step excitation when $P_2$ is given by Eq. (19), but this is not true for $P_1 > 4$. A more complicated excitation scheme would be required to excite the high-efficiency solution in this case [24]. The particular class of "easily accessible" steady-state solutions shown in Fig. 2(c) has a slightly increased domain of universal stability ($U_1 < 1$, $P_1 \leq 4/E_{ff}$, indicated by the region to the left of the dashed line in Fig. 2(c)) which extends beyond the region of monostability. In particular, universal stability is guaranteed in this set of solutions for all $P_1 < 4$.

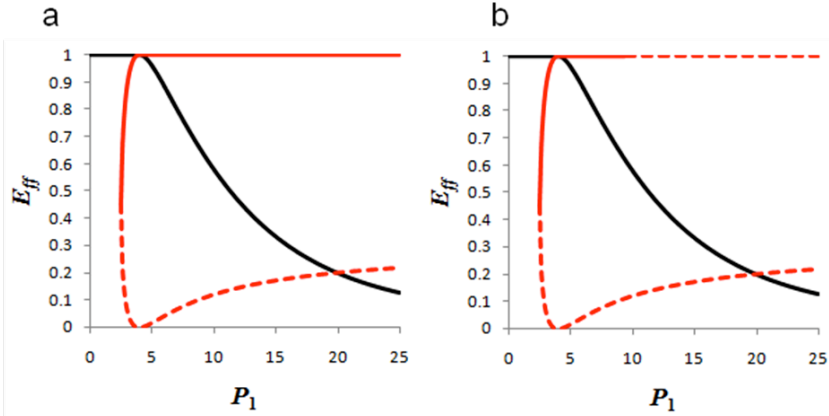

**Figure 3.** Plots showing the effect of geometry $\{r_2, r_T\}$ on the conversion efficiency and stability of steady-state solutions for $P_2 = (1 - P_1/4)^2$, with (**a**) $r_2 = r_T = 0.4$ and (**b**) $r_2 = r_T = 2$. The onset of multiple solutions occurs at $P_1 \approx 2.55$. Solid lines indicate stable conversion efficiencies; dashed lines indicate unstable conversion. The black line indicates the solution that is approached after a step excitation ($U_k$ ($T = 0$) = 0, $P_k$ ($T \geq 0$) = constant, $P_k(T < 0) = 0$) and red lines are use to denote all other solutions. While the conversion efficiency of all solutions is geometry-independent, note that the stability of quantum-limited efficiency changes with geometry ($r_2, r_T$). In (**a**) the high-efficiency solution is stable everywhere in the plot, but is unstable for $P_1 > \sim 10$ in (**b**) when $r_2$ and $r_T$ are increased from 0.4 to 2. For sufficiently large $r_2$ and $r_T$, the system approaches mono-stability for all $P_1$ (all solutions denoted in red become unstable).

## 5. Validity Limits:

We now consider which regions of the dimensionless parameter space are accessible for a given geometry ($Q_k$, $\omega_k$, $\beta$). Sufficient resonant enhancement of the density of optical modes at each of the three frequencies ($Q_k \gg 1$) is required in order to effectively suppress coupling to other frequencies (e.g. the sum frequency, $\omega_1 + \omega_2$). CMT relies on the slowly-varying amplitude approximation ($|da_k/dt| \ll |\omega_k a_k|$). This condition can be mapped into the dimensionless parameter space in the steady state, imposing the following conditions on the cavity Q-factors:

$$\frac{U_1}{2Q_2} \ll 1, \frac{U_2}{2Q_1} \ll 1 \qquad (26)$$

In the case that $E_{ff} = 1$, Eq. (26) reduces to,

$$P_1 \ll 8Q_2 \qquad (27)$$

When Eqs. (26-27) are not satisfied, it implies that the nonlinear terms in Eqs. (2-4) are sufficiently large that the perturbative treatment of the nonlinearity and the slowly varying amplitude approximation are no longer valid. A final consideration is that the energy in the cavity must be sufficiently below the threshold for the appearance of higher-order regimes or material damage. While Eqs. (26-27) ensure that second-order nonlinear effects remain in the perturbative limit, the nonlinear coupling strength (quantified by $\beta$) must also be sufficiently large for high-efficiency conversion to be achievable below the damage threshold. Efficiency can be expressed as a function of only the generalized energy in the idler cavity, $U_2$. Quantum-limited conversion requires $U_2 = 1$, and optimal total power conversion ($P_1 = 4$, $P_2 \rightarrow 0$) occurs when $U_k = 1$ for all three modes. When this occurs, the actual mode energies are given by $|a_k|^2 = Q_k/4Q_1Q_2Q_T|\beta|^2$. This energy must be below the material and frequency dependent threshold for the appearance of higher-order nonlinear effects or material damage (breakdown) in a specific geometry. Increasing the nonlinear mode-overlap, $\beta$, reduces the energy required for high-efficiency conversion.

## 6. Conclusion:

In this paper we have presented a general analysis of nonlinear second-order difference frequency generation (DFG) where all three waves correspond to resonant modes. We have shown that the equations governing the system can be scaled into a set of universal dimensionless parameters and that the steady-state solutions (corresponding to CW conversion) only depend on the generalized parameters, $P_1$ and $P_2$, which are directly proportional to the input powers of the pump and idler, respectively. As $P_1$ increases for any fixed $P_2$, a saddle-node bifurcation is reached. Below this point, there exists only one steady-state solution with universal stability, and above this point, there exist three solutions with geometry-dependent stabilities. In many cases, this results in bistability. We have demonstrated that CW conversion from pump to signal with quantum-limited efficiency is possible in such a system for *any* power of the pump wave, provided a specific input power at the idler frequency. In this case, the pump beam is perfectly depleted and the pump:signal conversion ratio is given by the frequency ratio. Quantum-limited conversion can occur with arbitrarily small idler power – optimizing total power conversion - when the pump power approaches the critical value. This power scales as the inverse of the product of the three cavity mode Q-factors, meaning that optimal conversion occurs at higher power with decreased Q-factors and vice-versa. This critical power also acts as a threshold below-which quantum-limited conversion is always stable and above-which the stability depends on geometry. We have shown that linear losses do not qualitatively change the dynamics, but do cause a reduction in conversion efficiency. The efficiency is maximized when theoretical Q-factors are kept sufficiently low that they are limited by coupling to the input/output port. We expect a similar universal framework to exist for second-order sum-frequency generation as well.

**Acknowledgements:**

We would like to thank Ken Kamrin, John Joannopoulos and Marin Soljacic for helpful discussions. IBB and MWM wish to acknowledge NSERC (Canada) for support from PGS-M and PDF fellowships. This work is supported through NSEC at Harvard, by the Army Research Office through the ISN under Contract No. W911NF-07-D-0004, and by US DOE Grant No. DE-FG02-97ER25308 (ARW).